\documentclass[
reprint,
 aps,
 amsmath,amssymb,
 prf,onecolumn,
]{revtex4-2}

\usepackage[utf8]{inputenc}
\usepackage{physics}
\usepackage{graphicx}
\usepackage{bm}
\usepackage[colorlinks=true, allcolors=blue]{hyperref}

\newcommand\We{\textrm{We}}
\newcommand\Oh{\textrm{Oh}}
\newcommand\Rey{\textrm{Re}}
\newcommand\Ca{\textrm{Ca}}

\usepackage{xcolor}

\begin{document}
\title{Deformation and stability of a gas bubble in a biaxial straining flow}
\author{Ali\'enor Rivi\`ere$^1$}
\email{alienor.riviere@epfl.ch}
\author{David Fabre$^2$}
\author{Jacques Magnaudet$^2$}
\author{François Gallaire$^1$}
\affiliation{$^1$ Laboratory of Fluid Mechanics and Instabilities, Ecole Polytechnique F\'ed\'erale de Lausanne, CH-1015 Lausanne, Switzerland}
\affiliation{$^2$ Univ. Toulouse; Toulouse INP, CNRS, IMFT (Institut de M\'ecanique des Fluides de Toulouse), F-31400 Toulouse, France}

\date{\today}

\begin{abstract}
Taking advantage of the recently developed L-ALE framework [Sierra-Ausin \textit{et al.}, Phys. Rev. Fluids {\bf{7}}, 113603 (2022)], we characterize the linear dynamics of an incompressible gas bubble immersed in a biaxial straining flow. We show that the system undergoes a saddle-node bifurcation with strongly different equilibrium shapes when varying the Ohnesorge number, $\Oh$, which compares viscous and capillary effects. Equilibrium shapes are found to be oblate for sufficiently large $\Oh$ while, counter-intuitively, they are prolate for low-enough $\Oh$. The bifurcation diagram is found to contain also two sets of disconnected branches that cannot be obtained by continuation starting from a spherical shape. One set corresponds to bubble shapes expected to be unstable, while the second set comprises a wide region exhibiting stable shapes that might be observed in practice. We then characterize the linear stability of the various branches. In addition to the unstable axisymmetric mode arising at the saddle-node bifurcation, two non-oscillating drift modes are also identified, together with a new unstable non-oscillating mode with azimuthal wave number $m=2$. This mode might be responsible for some type of bubble breakup observed in experiments.
\end{abstract}
\maketitle

\section{Introduction}
Reducing flow and geometry complexity frequently leads to a better understanding of the main physical ingredients controlling challenging problems. The breakup of drops and bubbles in turbulent flows is part of them.
Early attempts
showed that the dynamics of bubbles immersed in a uniaxial straining flow, characterized by one straining axis and two orthogonal compressional directions, can capture some of the important features of turbulent bubble breakup~\cite{revuelta2006,rodriguez2006,revuelta2008}. 
Building on this original intuition, recent studies correlated bubble deformation primarily with the largest strain direction, often neglecting the influence of the secondary strain \cite{masuk2021}. These works report a clear alignment of the bubble's major axis with the principal straining direction. However, the biaxial straining configuration, with one compressional axis and two stretching directions, is the most probable elementary local flow encountered in isotropic turbulence, as shown by the statistical distribution of the velocity-gradient invariants \cite{meneveau2011}. This observation remains valid across scales within the inertial range, as confirmed by coarse-graining analyses of the velocity-gradient tensor \cite{naso2005}. This raises the natural question of the stability of bubbles immersed in a biaxial straining flow, a configuration that has received considerably less attention than its uniaxial counterpart.

The dynamics of an immiscible drop or bubble freely suspended in a viscous fluid undergoing a uniaxial strain has attracted sustained attention since the pioneering experiments of Taylor using the four-roll-mill configuration~\cite{taylor1934}. In the absence of inertia, the drop shape results from the competition between viscous and capillary stresses, quantified by the capillary number, $\Ca$. Beyond a critical value  
$\Ca=\Ca_c$, a saddle-node bifurcation leads to drop breakup. In contrast, for a bubble with zero inner viscosity, this critical capillary number tends to infinity \cite[fig. 4.8]{gallino2018thesis}, pointing to a seemingly singular limit under Stokes flow conditions.

However, inertial effects can no longer be neglected for gas bubbles immersed in turbulent flows.
The bubble shape is then determined by a balance between pressure and viscous stresses, which promote deformation, and capillary stresses, which oppose it. The relevant governing parameters then become the Weber number, $\We$, and the Reynolds number, $\Rey$, characterizing the ratios of inertial to capillary and viscous forces, respectively.
Equivalently, one may use the Ohnesorge number $\Oh=\We^{1/2}\Rey^{-1}$ characterizing the ratio of viscous and capillary stresses instead of $\Rey$.
In uniaxial straining flows, Acrivos and Lo \cite{acrivos1978} showed that no steady bubble shape can exist beyond a critical Weber number 
$\We_c$ increasing as $\Rey^{3/4}$.
Subsequent studies by Miksis \cite{miksis1981}, Ryskin \& Leal~\cite{ryskin1984},  Kang \& Leal~\cite{kang1987uniaxial,kang1988shape} and Rivi\`ere \textit{et al.}~\cite{riviere2023}, who investigated the time-dependent bubble evolution from arbitrary initial conditions, further mapped the bifurcation structure. Recently, Sierra-Ausin \textit{et al.} \cite{sierra-Ausin2022} revisited this classical problem by performing a global linear stability analysis. Since the flow domain is time-dependent owing to the unknown bubble shape,  they developed a Linearized Arbitrary Lagrangian-Eulerian (L-ALE) approach \cite{bonnefis2019thesis,sierra-Ausin2022,bonnefis2024} which enables the governing equations and boundary conditions to be rigorously expanded on a time-independent reference domain.

For perturbations preserving the basic flow symmetries, the bifurcation diagram consists of one stable branch and one unstable branch of steady states, connected by a saddle-node bifurcation. Along these branches, several unstable global modes may develop. An oscillatory mode that becomes neutrally stable in the inviscid limit is found along the stable branch while, unsurprisingly, a non-oscillatory unstable mode is detected along the unstable branch. When the basic flow symmetries are relaxed, i.e., the bubble is not constrained to remain at the stagnation point of the base flow, two extra non-oscillatory unstable modes arise, corresponding to a drift of the bubble centroid away from the stagnation point. One of them represents a translation along the extensional axis while, more counter-intuitively, the other corresponds to a drift within the compressional plane, i.e., against the undisturbed flow.

Apart from the pioneering work of Kang \& Leal \cite{kang1989biaxial}, who investigated the bubble equilibrium shapes and deformation dynamics in a biaxial flow \textit{via} a time-marching solution of the vorticity equation (an approach unable to capture unstable equilibria), most existing analyses of drops or bubbles in biaxial straining flows have been confined to the creeping-flow limit.
Zabarankin \textit{et al.} \cite{zabarankin2013} conducted a thorough study of equilibrium drop shapes under Stokes flow conditions for various viscosity ratios, while Malik \textit{et al.} \cite{malik2020} incorporated rotation effects in the same limit. A recent extension by Malik \textit{et al.} \cite{malik2024} examined both uniaxial and biaxial flows with rotation, connecting the dimpled, simply connected drop shapes to the toroidal morphologies described by Zarabankin \textit{et al.} 
\cite{zabarankin2015} and Ee \textit{et al.} \cite{ee2018} (see also \cite{lavrenteva2021} for the connection of dimples and tori in the absence of rotation).

In the present study, we build upon the L-ALE framework of \cite{sierra-Ausin2022} to obtain the complete bifurcation diagram and perform a linear stability analysis of bubbles in biaxial straining flows, accounting for arbitrary inertial effects. Section \ref{sec:NumMethod} introduces the physical configuration and summarizes the numerical methodology. Section \ref{sec:eqpos} explores the equilibrium states, revealing a saddle-node bifurcation at a critical $\Oh$-dependent 
Weber number, and markedly different phenomenologies when the Ohnesorge number is varied. Section \ref{sec:LSA} presents the stability analysis of the various branches. We first characterize the least stable and most unstable axisymmetric modes, then identify two drift modes analogous to those found in uniaxial straining flows \cite{sierra-Ausin2022}. Finally, we report the emergence of a new unstable mode with azimuthal wavenumber 
 $m=2$ which may underlie some of the binary breakup events observed in turbulent flows.

\section{Methodology}
\subsection{Problem formulation}
\label{sec:NumMethod}

\begin{figure}
    \centering
    \includegraphics{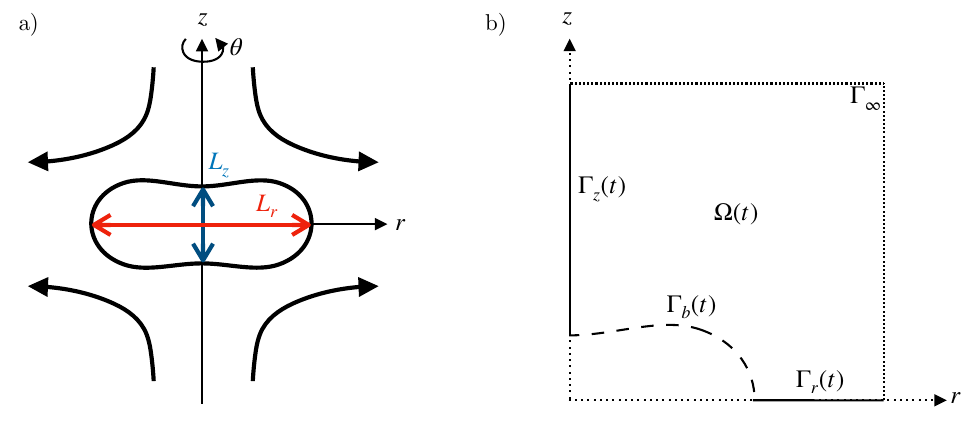}
    \caption{(a) Schematic representation of a bubble in a biaxial straining flow in a $(r, \theta,z)$ cylindrical coordinate system. Black arrows schematize the streamlines of the undisturbed flow. Lengths $L_r$ and $L_z$ quantify the bubble lengths in the $z=0$ plane and along the $r=0$ axis, respectively. (b) Sketch of the numerical domain $\Omega$, not to scale. $\Gamma_b(t)$ denotes the time-dependent interface position, while $\Gamma_r(t)$ and $\Gamma_z(t)$ are the boundaries of the domain along the $z=0$ symmetry plane and the $r=0$ symmetry axis, respectively. Finally, $\Gamma_\infty$ represents the outer boundary that closes the computational domain.}
    \label{fig:scheme}
\end{figure}
We consider an incompressible gas bubble of volume $\cal V$, equivalent diameter $d = \sqrt[3]{6 \cal V/\pi}$, filled with a gas having negligible viscosity and density. The bubble is immersed in a liquid with density $\rho$, kinematic viscosity $\nu$ and dynamic viscosity $\mu = \rho \nu$. The surface tension acting at the gas-liquid interface is $\gamma$. As schematized in Fig.~\ref{fig:scheme}(a), the bubble is placed at the stagnation point of an incompressible axisymmetric biaxial straining flow of velocity $\vb{u}_{BSF} = -S z\vb{e_z} +  \frac{1}{2} Sr \vb{e_r}$, and pressure $P_{BSF} = -S^2 (\frac{1}{8}r^2 + \frac{1}{2}z^2)$, characterized by a strain-rate $S\geq 0$, with $(r, z)$ denoting the radial and axial cylindrical coordinates.
The bubble dynamics is controlled by the balance between inertial, capillary and viscous forces, from which two independent dimensionless numbers can be built. 
In the following, we investigate the bubble dynamics in the phase space built on the Weber number, $\We = \rho U^2 d/\gamma = \rho S^2 d^3/(4\gamma)$, with the characteristic velocity scale $U = S d/2$, and the Ohnesorge number $\Oh = \mu/\sqrt{\rho \gamma d}$. In this setting, the Weber number, which compares inertial and capillary forces, measures the bubble deformability, while the Ohnesorge number defines the bubble size with respect to the viscous-capillary length scale $\mu^2/(\rho\gamma)$.
Other choices of dimensionless groups are of course possible.
In particular, one can define a Reynolds number $\Rey = \We^{1/2}\Oh^{-1} = S d^2/2\nu$ from the balance between inertial and viscous forces.
In the limit of large $\Oh$, inertial forces become negligible. We will therefore compare our results obtained in that limit with creeping-flow results from the literature, which are usually discussed in terms of the capillary number $\Ca = \We /\Rey = \mu U/\gamma = \mu Sd/(2\gamma)$, which compares viscous and capillary forces.

In the following, we quantify bubble deformations using the shape parameter $\Psi = (L_z -L_r)/(L_z + L_r)$, where $L_z$ and $L_r$ denote the two bubble lengths measured along the flow axes as defined in Fig.~\ref{fig:scheme}(a). This metric, which was first introduced by Taylor \cite{taylor1934}, is a practical alternative to the aspect ratio $\chi$ used in \cite{sierra-Ausin2022} which is the ratio of the major and minor bubble lengths along the axes of the base flow. The advantage of the present choice is that $\Psi$ is bounded in the interval $[-1, 1]$, the case $\Psi =0$ corresponding to a sphere, while $\Psi > 0$ and $\Psi < 0$ correspond to elongated (prolate) and flattened (oblate) shapes, respectively. 

The problem is governed by the axisymmetric Navier-Stokes equations on the time-dependent domain $\Omega(t)$ sketched in Fig.~\ref{fig:scheme}(b),
\begin{align}
\rho \left [ \partial_{t, \Omega(t)}\vb{u} + \vb{u} \cdot \vb{\nabla}_{\Omega(t)} \vb{u} \right ] &= \vb{\nabla}_{\Omega(t)} \cdot \vb{\Sigma}_{\Omega(t)} & \text{in } \Omega(t)\,,
\label{eq:GoverningEq1}
\\
\vb{\nabla}_{\Omega(t)} \cdot \vb{u} & = 0  & \text{in } \Omega(t)\,,
 \label{eq:GoverningEq0} \\
\partial_{t, \Omega(t)} \eta &= \vb{u}\cdot \vb{n}  & \text{on } \Gamma_b(t)\,,
\label{eq:GoverningEq2} \\
\vb{\Sigma}_{\Omega(t)} \cdot \vb{n} &= \left( -p_b + \gamma \kappa \right ) \vb{n} &  \text{on }\Gamma_b(t)\,,
 \label{eq:GoverningEq3}
\end{align}
where $\Gamma_b(t)$ denotes the instantaneous interface position.
The above governing equations are supplemented with appropriate boundary conditions on the symmetry axis $r=0$ ($\Gamma_z$), the symmetry plane $z=0$ ($\Gamma_r$) and in the far field ($\Gamma_\infty$). In addition, the solution of \eqref{eq:GoverningEq1}-\eqref{eq:GoverningEq3} must ensure that the bubble volume does not change over time, which adds an integral constraint over $\Gamma_b(t)$. In (\ref{eq:GoverningEq1})-(\ref{eq:GoverningEq3}), the subscript $_{\Omega(t)}$ is used to stress the fact that time and space derivatives are to be evaluated in the time-dependent domain $\Omega(t)$.
Lengths are expressed in units of the initial bubble radius $R= d/2$. Times are in units of the capillary timescale $\sqrt{\rho R^3/\gamma}$.
The kinematic boundary condition (\ref{eq:GoverningEq2}) implies that, at any location $\bf{x}$ on the interface, the time derivative of the interface displacement $\eta({\bf x},t)$ in the normal direction (with unit vector ${\bf n}({\bf x},t)$) must coincide with the normal component ${\bf u}\cdot{\bf n}$ of the local fluid velocity ${\bf u}({\bf x},t)$. The dynamic boundary condition (\ref{eq:GoverningEq3})
is expressed in terms of the stress tensor in the carrier fluid, ${\bf \Sigma}_{\Omega}({\bf u},p) = - p {\bf {I}} + 2\mu {\bf D}_{\Omega}({\bf u})$, with $p$, ${\bf I}$ and ${\bf D}_{\Omega}({\bf u})$ the pressure, unit tensor and  strain-rate tensor, respectively. The normal projection of (\ref{eq:GoverningEq3}) expresses the fact that the normal stress ${\bf n}\cdot{\bf \Sigma}_{\Omega}\cdot{\bf n}$ balances the difference between the uniform pressure $p_b(t)$ inside the bubble and the local capillary pressure $\gamma \kappa{\bf n}$, with $\kappa({\bf  x},t)=\nabla\cdot{\bf n}({\bf x},t)$ the local mean curvature of the interface. 
Last, the tangential projection of (\ref{eq:GoverningEq3}) yields the shear-free condition ${\bf n}\times({\bf \Sigma}_{\Omega}\cdot {\bf n}) ={\bf 0}$ which holds if the interface is free of any contamination. 

\subsection{Numerical solution}
\label{numsol}

The numerical solution of the problem is obtained in essentially the same way as in 
\cite{sierra-Ausin2022} and relies upon the L-ALE framework developed in \cite{bonnefis2019thesis}. The method handles the deformation of the fluid domain by introducing an arbitrary displacement field $\bf \xi({\bf x},t)$ which extends the normal displacement of the free surface $\eta$ throughout the bulk domain $ \Omega(t)$. We refer the reader to  Appendix A of \cite{sierra-Ausin2022} for a detailed presentation of the method and only give here a synthetic overview.

The first stage consists in computing stationary equilibrium shapes and the associated stationary flows. This is achieved by using an iterative Newton method which directly solves the stationary version of (\ref{eq:GoverningEq1})-(\ref{eq:GoverningEq3}) (i.e., the time derivatives are dropped), starting from a guess approximate solution typically corresponding to an equilibrium state obtained for a previous set of the control parameters $\Oh-\We$. A pseudo-arc-length continuation method is employed, allowing equilibrium solutions in the $\We-\Psi$ plane to be followed even in the presence of a saddle-node bifurcation connecting two branches of the bifurcation diagram. Thanks to the symmetries of the problem, the solution is computed in the quadrant $r>0$, $z>0$ only, with $(r,z) \in [0, 10]^2$ (see Fig.~\ref{fig:scheme}b); the other quadrants may then be reconstructed by symmetry for plotting purposes. For the sake of simplicity, $r$ will be considered negative in the left two quadrants in figures displaying the entire flow field about the bubble.

In a second stage, these equilibrium states are used as base states to perform a global stability analysis, which consists in examining the evolution of disturbances with a prescribed eigenmode form. 
In the present problem, the base configuration exhibits an axial symmetry about the $z$-axis. It is thus relevant to consider velocity, pressure and surface deformation disturbances with the form $e^{i (m \theta-\lambda t)}$, where $\theta$ is the azimuthal angle in the cylindrical coordinate system, $m$ is the corresponding wavenumber and $\lambda=\omega+i\sigma$ is the complex eigenvalue. 
Unstable eigenmodes satisfying $\mathcal{I}m(\lambda)=\sigma > 0$ can be classified as stationary (s) if
$\mathcal{R}e(\lambda)=\omega=0$ or oscillating (o) if $\mathcal{R}e(\lambda)\neq0$. They can also be classified as symmetric (S) or antisymmetric (A) with
respect to the plane $z = 0$. In what follows, we classify the modes using a nomenclature that summarizes their three characteristic properties, starting with their azimuthal wavenumber $m$. 
For instance  a `(0 - S) (s)' mode is axisymmetric $(m = 0)$, symmetric with respect to the plane $z=0$, and stationary. Note that the `stationary' terminology refers to the non-oscillatory nature of the mode but implies by no means that its amplitude does not change over time, which is the case whenever $\sigma\neq0$.

The numerical solution is obtained by combining a finite-element discretization for the bulk unknowns (velocity $\bf u$, pressure $p$, displacement field $\bf \xi$) and a Fourier discretization for the surface unknowns (displacement $\eta$ and curvature $\kappa$). This enables the construction of the Jacobian operator of the problem in a block-matrix form. This operator is used in both the Newton method for the base-flow calculation and the eigenvalue computation for the stability analysis. All steps are implemented within the {\sc FreeFem} software \cite{hecht2012,freefem}, and eigenvalues are computed using a Krylov method thanks to the SLEPc library. Note that mesh adaptation is used along the iteration process, and this ingredient proves to be essential when the continuation method leads to strongly deformed shapes. 
All codes used here are interfaced thanks to the StabFem interface \cite{fabre2018}. Sample codes reproducing the key steps of the calculations may be found on the StabFem website \cite{stabfem}.

\section{ Equilibrium shapes }
\label{sec:eqpos}

\begin{figure}
    \centering
    \includegraphics{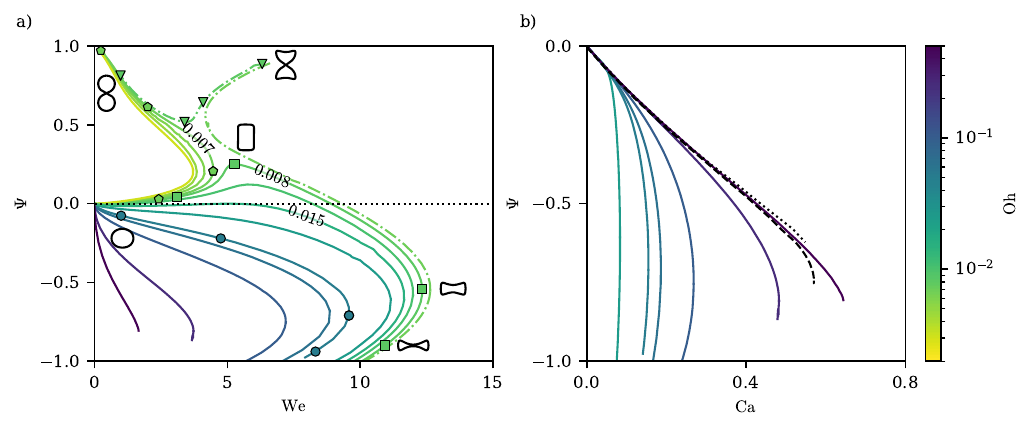}
    \caption{Equilibrium shapes. (a) Variations of the stationary bubble shape with $\We$ for various $\Oh \in \{3.10^{-3}, 5.10^{-3}, 6\cdot 10^{-3}, 7\cdot 10^{-3},  8\cdot 10^{-3}, 0.01,0.015,  0.025, 0.05, 0.0625 ,  0.1, 0.25, 0.5\}$ (color-coded, from yellow to black). Solid curves correspond to states computed by continuation, starting from a spherical shape $\Psi = 0$ \textendash{} $\We = 0$, while dash-dotted lines correspond to disconnected branches. Values of $\Oh$ corresponding to transitions between the regimes discussed in Sec. \ref{sec:eqpos} are indicated on the corresponding curves. 
    Symbols along the curves correspond to the parameters for which the steady solutions are represented on figures~\ref{fig:BF_Oh0v05}-\ref{fig:BF_otherbranchesOh0v008}. Some typical shapes are also shown. (b) Variations of the shape factor $\Psi$ for the six largest $\Oh$ as a function of the capillary number $\Ca$. As $\Oh$ increases, branches converge toward the creeping-flow solutions of \cite{kang1989biaxial} (black dotted line) and \cite{zabarankin2013} (black dashed line).}
    \label{fig:phase diagram}
\end{figure}

In this section, we investigate the equilibrium states of the system, computed thanks to the continuation method outlined above. Figure~\ref{fig:phase diagram}(a) provides a general view of the bubble shapes in a diagram in which the deformation parameter $\Psi$ is shown as a function of the Weber number $\We$, for thirteen values of the Ohnesorge number $\Oh$ ranging from $\Oh=0.003$ to $0.5$. Two families of curves are seen in this figure. The first ones (depicted by solid lines regardless of their stability properties which will be addressed in the next section) are called {\em main branches} since they can be computed by continuation,
starting from an unstrained spherical bubble ($\Psi = 0$, $\We = 0$). The second series of curves, depicted with dash-dotted lines, consists of {\em disconnected branches} which can only be obtained using a different continuation process, as will be explained later. 
The figure shows that strikingly different shapes are obtained for viscosity-dominated (large $\Oh$) and capillary-dominated (small $\Oh$) configurations, with a sharp transition occurring in the intermediate $\Oh$-range. In the next subsections we will comment on the trends observed in the successive regimes revealed by this diagram, and depict the equilibrium shapes encountered when moving along the curves corresponding to a fixed $\Oh$.

\subsection{Viscosity-dominated range $(\Oh \geq 0.015)$}

\begin{figure}
    \centering
    \includegraphics{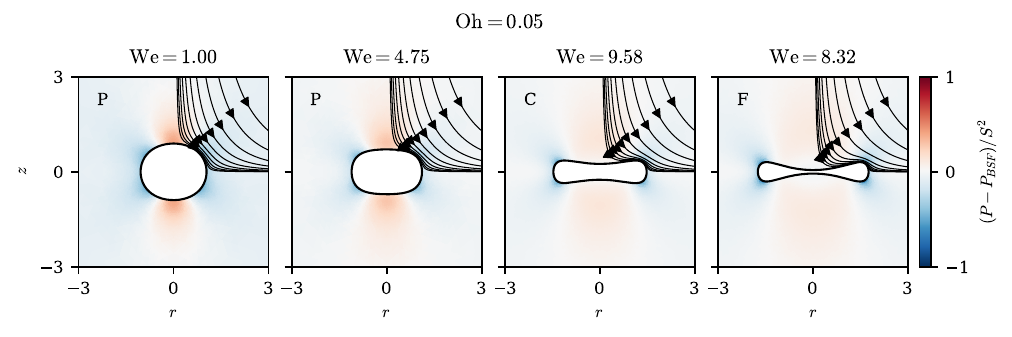}
    \caption{Example of steady solutions obtained at $\Oh=0.05$, shown with bullets in Fig.~\ref{fig:phase diagram}(a). The upper right quadrant is computed numerically; the other three quadrants are reconstructed by symmetry. 
  The bubble-induced pressure disturbance with respect to the base flow, $(P - P_{BSF})/S^2$, is color-coded. Some streamlines are depicted with thin black lines and arrows. As $\We$ increases on the primary branch (P), the bubble becomes more oblate. At the critical point (C), $\We = 9.58$, the bubble becomes concave in the vicinity of the symmetry axis, a trend that further increases on the folded branch (F), until $L_z$ vanishes.}
    \label{fig:BF_Oh0v05}
\end{figure}

We first examine the equilibrium shapes along the main branches computed for cases with $\Oh\geq0.015$ (plain lines with the darkest colors in Fig.~\ref{fig:phase diagram}(a)). In this range, all computed equilibrium states lie on the $\Psi<0$ side of the diagram, corresponding to flattened shapes. Moreover, all curves undergo a turn towards the left after passing through a maximum Weber number. This means that, when increasing $\We$ starting from zero, two branches of solutions exist, which eventually merge at the point corresponding to the maximum $\We$. Throughout the rest of the paper, branches of solutions emerging from the spherical shape will be referred to as \textit{primary} and denoted (P), whereas the second branch of solutions connected to the primary branch will be referred to as \textit{folded} and denoted (F). The existence of a primary branch and a folded branch is
 symptomatic of a saddle-node bifurcation, and it can thus be anticipated that the folded branch is always unstable. This type of bifurcation was already observed in both biaxial Stokes flow \cite{zabarankin2013,gallino2018} and uniaxial straining flows \cite{kang1987uniaxial,miksis1981,sierra-Ausin2022}. 


Let us consider in more detail the case $\Oh = 0.05$ which is characteristic of the `large-$\Oh$' range. Figure~\ref{fig:BF_Oh0v05} displays the bubble shape and steady flow structure at four points along the main branch (corresponding to bullets in Fig.~\ref{fig:phase diagram}(a)). Starting from the spherical shape on the primary branch (P) and increasing the Weber number towards $\We=1$ and then $\We=4.75$, the bubble shape progressively flattens. 
At the critical point $\We = \We_c= 9.58$ (C), the curvature changes sign on the symmetry axis. When moving further along the folded branch (F) (which can only be achieved using arc-length continuation), the negative curvature on the symmetry axis and the extent of the concave region increase as $\We_c-\We$ increases, as the bubble shape corresponding to $\We= 8.32$ confirms. Continuing along the folded branch, the upper and lower sides of the bubble eventually touch each other for $\We \approx 8$, yielding $\Psi= -1$. Beyond this point, one can expect the existence of toroidal shapes, but obviously these ones cannot be obtained through the present continuation method.


For the largest values of $\Oh$ (darkest curves in the figure), the location of the turning point moves towards smaller $\We$. In this range, 
inertia becomes subdominant and the equilibrium solutions can be expected to 
converge towards those obtained in the creeping-flow limit. It is thus more appropriate to describe these results in terms of the capillary number $\Ca = \We /\Rey$.
Figure~\ref{fig:phase diagram}(b) compares the finite-inertia datasets obtained for the six highest $\Oh$ ($\Oh > 0.015$ (darkest curves in Fig. \ref{fig:phase diagram}(a)) with the equilibrium solutions of Kang and Leal~\cite{kang1989biaxial} and Zabarankin \textit{et al.}~\cite{zabarankin2013} corresponding to the creeping-flow regime. 
For small enough deformations, the deformation factor decreases linearly with $\Ca$ following Taylor's prediction \cite{taylor1934} 
$\Psi = -3 \Ca + O(\Ca^2)$, obtained through a first-order domain perturbation analysis of a drop deforming in a linear shear flow (see also \cite{leal1992}).
At the largest Ohnesorge number, $\Oh = 0.5$, we obtain equilibrium shapes up to $\Ca_c\approx0.643$, which is larger than the critical $\Ca$ reported in \cite{zabarankin2013} in the creeping-flow limit, $\Ca_c= 0.571$.
Since $\Rey\geq2$ when $\Ca \geq 0.5$ for $\Oh = 0.5$, we hypothesize that in this range of parameters inertia still plays a role, allowing larger bubble deformations than in the creeping-flow limit. 

Note that for the largest two Ohnesorge numbers ($\Oh = 0.5$ and $\Oh = 0.25$), despite our efforts, we were not able to follow the folded branches beyond the saddle-node point and towards the limit $\Psi = -1$. In this range, spurious oscillations appear close to the symmetry axis, leading to a divergence of the Newton solver. Yet, we do not expect these branches to present specific behaviors.
\subsection{ Transition range $(0.008 \leq \Oh < 0.015)$ }
\begin{figure}
    \centering
    \includegraphics{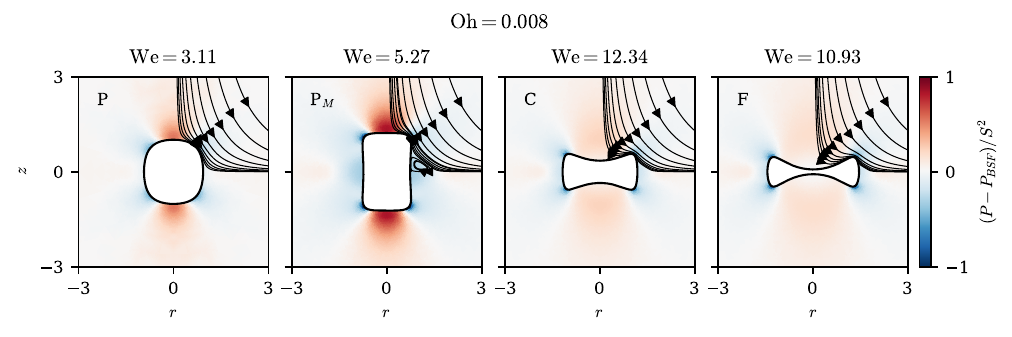}
    \caption{Example of steady solutions obtained for $\Oh=0.008$ (squares in Fig.~\ref{fig:phase diagram}(a)). For details see the caption in Fig.~\ref{fig:BF_Oh0v05}. On the primary branch, the bubble elongates along the axial direction, forming a prolate cylinder. At $\We=5.27$ (P$_M$), the extension is maximal and then decreases, yielding oblate bubble shapes with relatively sharp edges along the folded branch (F).}
    \label{fig:BF_Oh0v008}
\end{figure}

We now return to Fig.~\ref{fig:phase diagram}(a) and consider the trends observed when $\Oh$ is decreased below $0.015$. A first striking feature is that, for moderate $\We$, shapes with $\Psi>0$ are obtained. This means that, rather counter-intuitively, a biaxial strain of moderate intensity now leads to prolate shapes, just like a uniaxial strain, instead of oblate shapes.

To describe this feature in more detail, we focus on the case $\Oh = 0.008$. Figure~\ref{fig:BF_Oh0v008} displays some equilibrium shapes obtained when moving along the main branch (for parameters corresponding to squares in Fig.~\ref{fig:phase diagram}(a)).
For $\We = 3.11$ (P), a slightly prolate shape is obtained.  
The maximum elongation is reached at $\We = 5.27$ (P$_M$), for which the bubble shape resembles a cylinder with rather sharp edges. Note that, in this range of parameters, toroidal recirculation regions form at the bubble periphery, contributing to decrease the (negative) axial curvature near the symmetry plane.
The cylindrical shapes reported here were first observed by Kang and Leal \cite{kang1989biaxial} for both $\Rey \geq 200$ and in potential flow simulations.
They arise when inertial pressure forces dominate over viscous forces at the bubble interface.
Pressure gradients being four times larger along the axial direction than along the radial one due to the pressure distribution $P_{BSF}(r,z)$ in the base flow, the bubble shape extends along the symmetry axis, yielding a prolate shape.

As $\We$ is further increased along the main branch, the cylindrical 
shape is maintained, but progressively becomes more compressed in the top and bottom regions, eventually turning back to an oblate shape ($\Psi<0$) and eventually a disk-like shape for $\We \gtrsim 8$. Increasing $\We$ again up to the turning point (C), a concave region appears along the axis, as previously observed at higher $\Oh$. When moving further along the folded branch (F), the bubble approaches again a toroidal shape with $\Psi \approx -1$. However, its edges are much sharper than observed previously at larger $\Oh$.
Consequently, the deformation parameter at the turning point decreases in absolute value with decreasing $\Oh$ (ranging from -0.8 at $\Oh =0.5$, to -0.54 at $\Oh = 0.008$), as deformations in the corresponding $\We$-range tend to squeeze the equatorial region of the bubble.

\subsection{Inertia-dominated range $(\Oh\leq0.007)$ }
\begin{figure}
    \centering
    \includegraphics{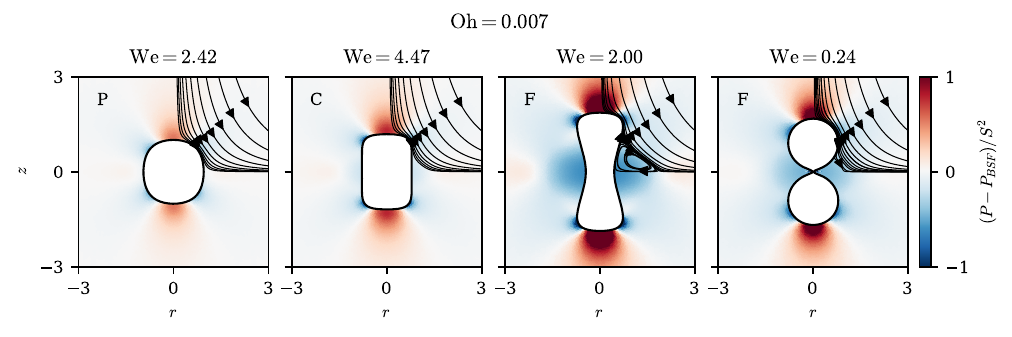}
    \caption{Example of steady solutions obtained for $\Oh=0.007$ (pentagons in Fig.~\ref{fig:phase diagram}(a)). For details see the caption of Fig.~\ref{fig:BF_Oh0v05}. At this $\Oh$, the bubble exhibits a prolate shape on the primary branch (P), turns approximately into a cylinder at the critical point (C), and becomes increasingly squeezed in the symmetry plane $z=0$ along the folded branch (F), until it completely necks down and forms a dumbbell.}
    \label{fig:BF_Oh0v007}
\end{figure}

When the Ohnesorge is further decreased to $0.007$ and below, an even more radical transition is observed. To illustrate this, Fig.~\ref{fig:BF_Oh0v007} displays shapes obtained with $\Oh = 0.007$ when continuously moving along the main branch (for $\We$-values corresponding to pentagons in Fig.~\ref{fig:phase diagram}(a)). For small $\We$, the deformation parameter $\Psi$ becomes positive and increases with the Weber number, indicating increasingly prolate shapes (P). Then, unlike for $\Oh=0.008$, the curve reaches a maximum $\We$ at a turning point (C), where the bubble exhibits a cylindrical shape with sharp edges. Beyond this turning point, the branch can be continued using arc-length continuation, revealing a presumably unstable folded branch turning back towards decreasing $\We$. Along this branch, the bubble shape is characterized by the occurrence of a neck in the symmetry plane  $z=0$ (F). The branch can be continued down to $\We \approx 0$, leading to a dumbbell shape made of two almost spherical bubbles connected by a narrow neck (F), which corresponds to $L_r \approx 0$, hence $\Psi \approx 1$.
Note that this limit shape is no surprise, since a shape consisting of two spherical bubbles touching each other is indeed a valid solution of the static problem for $\We = 0$.

\subsection{Disconnected branches}

\begin{figure}
    \centering
    \includegraphics{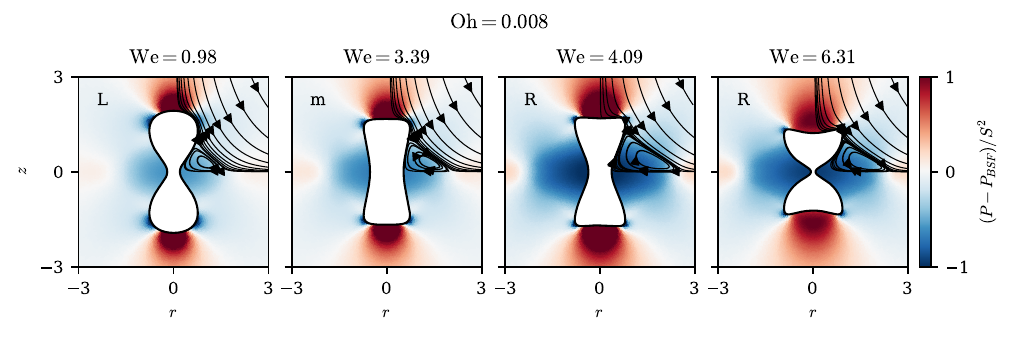}
    \caption{Example of steady solutions obtained for $\Oh=0.008$ on the disconnected V-shaped branch (triangles in Fig.~\ref{fig:phase diagram}(a)). For details see the caption of Fig.~\ref{fig:BF_Oh0v05}.
    Starting from the two-bubble solution (L), the neck reaches a maximum radius and, henceforth, $\Psi$ reaches a minimum at $\We = 3.39$ (m). For larger $\We$, the bubble adopts an hourglass shape (R).}
    \label{fig:BF_otherbranchesOh0v008}
\end{figure}

The sharp transition encountered for $\Oh \in [0.007, 0.008]$ suggests the existence of other branches of solutions, which are not continuously connected with the spherical shape. As another hint pointing towards the existence of such disconnected branches, one may argue that the limit case $\We \approx 0$ \textendash{} $\Psi \approx 1$ in which the bubble exhibits a dumbbell shape made of two spheres touching each other is a relevant limit of the problem whatever  $\Oh$. This remark suggests a practical method to reach such disconnected branches. Namely, we initialize the Newton algorithm with a steady solution at $\Oh = 0.007$ on the folded branch, and increase the value of Oh to 0.008. By doing so, we obtain a new solution that effectively belongs to a disconnected V-shaped branch and can be followed using arc-length continuation. This disconnected branch is plotted with a dash-dotted line in Fig.~\ref{fig:phase diagram}(a). On this branch, starting from $\Psi \approx 1$ at $\We \approx 0$, $\Psi$ first decreases, reaches a minimum value $\Psi = 0.52$ at $\We = 3.39$ and then re-increases, until  returning to $\Psi\approx 1$ at a finite $\We$. 

Figure~\ref{fig:BF_otherbranchesOh0v008} displays four bubble shapes computed along this V-shaped branch, corresponding to the triangular symbols in Fig.~\ref{fig:phase diagram}(a). From two approximately spherical bubbles connected by a neck at $\We = 0.98$ (L), on the left branch of the 'V', the bubble shape becomes more angular and nearly cylindrical as $\Psi$ reaches its minimum for $\We = 3.39$ (m). Moving further toward the right side of the V-shaped branch, the bubble adopts an hourglass shape (R) with both a narrow neck on the symmetry plane $z=0$ and concave parts away from this plane (top and bottom regions in the figure). This shape eventually leads to two disconnected concave disks for large enough $\We$.
We anticipate that these exotic bubble shapes are strongly unstable and cannot be observed in experiments. 

A second family of disconnected branches exists for $\Oh = 0.007$ and below. This branch cannot be captured by a continuation approach at fixed $\Oh$, but it can be reached by starting from a cylindrical shape at $\We \approx 7$\textendash{} $\Psi \approx 0.2$\textendash{} $\Oh=0.008$  and gradually decreasing $\Oh$ down to 0.007.  
The S-shaped disconnected branch obtained in this way, also plotted with a dash-dotted line in Fig.~\ref{fig:phase diagram}(a), spans all $\Psi$-values and bubbles shapes, from hourglass to cylinders, disks and finally tori. 
While this branch has never been described, Kang and Leal~\cite{kang1989biaxial} conjectured its existence. Starting from a spherical bubble at $\Rey= 400$, they gradually increased $\We$ up to a critical value ($\We =5$) above which no steady solution was found. Yet, starting from an equilibrium solution at $\Rey = 200$ \textendash{} $\We = 12$ and increasing gradually $\Rey$, they found new equilibrium solutions at $\Rey = 400$ for $\We \geq 7$.
They concluded that there is a limit point separating these two sets of solutions, a conclusion corroborated by present results.

\section{Linear stability}
\label{sec:LSA}
\begin{figure}
    \centering
    \includegraphics{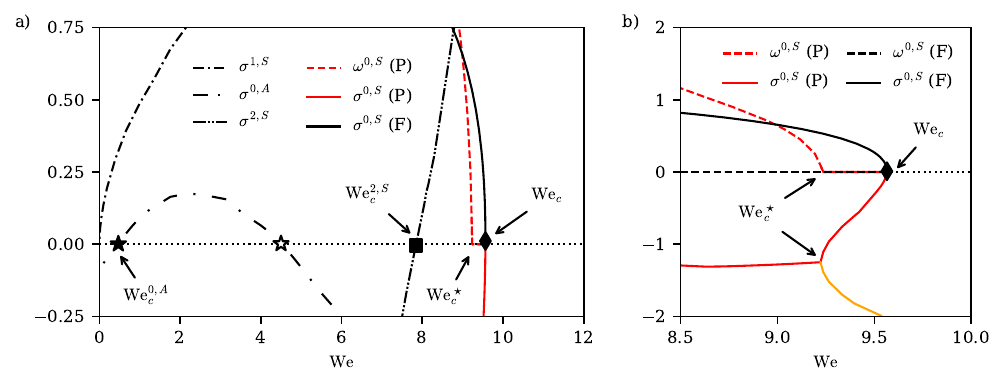}
    \caption{Emergence of unstable modes at $\Oh = 0.05$: (a) along the main branch and (b) enlargement close to the turning point. (a) Growth rates associated with the successive stationary modes at $\Oh = 0.05$. At $\We =0$, a (1 - S) (s) drift mode grows, followed by a (0 - A) (s) drift mode which is unstable in the finite interval bounded by the black and white stars. Further increasing $\We$, a (2 - S) (s) mode associated with non-axisymmetric deformations becomes unstable at the location of the black square. 
    Subsequently, as best illustrated in plot (b),  
    at $\We=\We_c^\star$, the frequency  $\omega^{0,S}$ of a pair of complex-conjugate stable oscillating modes with symmetry (0-S) becomes null, and two stable non-oscillating modes emerge, the (negative) growth rates of which are depicted by the red and orange lines. At the turning point corresponding to the critical Weber number $\We_c$ (diamond), one of them becomes unstable. Its growth rate increases along the unstable folded branch (solid black line).}
    \label{fig:summary_unstab}
\end{figure}

We now investigate the linear stability of the main branches of the bifurcation diagram.
The stability of the disconnected branches is more challenging to handle numerically, due to grid deformations, but can be inferred from that of the set of main branches.
We start by investigating modes (0 - S) (following the nomenclature defined in Sec. \ref{numsol}) that preserve the symmetries of the base flow, before discussing modes (1 - S) breaking the axial symmetry or modes (0 - S) that break the planar symmetry with respect to the plane $z=0$. Lastly, we describe higher-order modes ($m$ - S) and ($m$ - A) with $m>1$.
A summary of the successive instabilities can already be found in Fig.~\ref{fig:summary_unstab} for $\Oh = 0.05$ and will be described in detail in this section.

\subsection{Symmetry-preserving modes}
We first focus on the symmetry-preserving modes of (0 - S) type, for which the bubble centroid is held fixed at the stagnation point.
Similar to what has been observed in uniaxial straining flows, starting from a sphere, the base flow is stable with respect to axisymmetric disturbances until the turning point corresponding to a saddle-node bifurcation is reached at $\We=\We_c$.
At Weber numbers significantly lower than $\We_c$, the least stable modes are a pair of complex-conjugate oscillating modes with an almost constant damping rate $\sigma^{0, S}$ and a frequency $\omega^{0,S}$ that decreases with increasing $\We$ (dashed red line in Fig.~\ref{fig:summary_unstab}).
This frequency vanishes at a Weber number $\We_c^\star < \We_c$, with the interval $\We_c - \We_c^\star $ expected to shrink as $\Oh \to 0$, similar to what has been reported for uniaxial straining flows.
Two stationary modes emerge at $\We\geq\We_c^\star$, with growth rates diverging symmetrically from $\sigma^{0, S}$ (solid red and orange lines in Fig.~\ref{fig:summary_unstab}(b)). The one with the smallest damping rate eventually starts to grow at the turning point. Therefore, a stationary, i.e., not oscillating, (0 - S) (s) mode becomes unstable along the folded branch, with its growth rate increasing with the distance $\We_c$ - $\We$ (solid black line).

\begin{figure}
    \centering
    \includegraphics{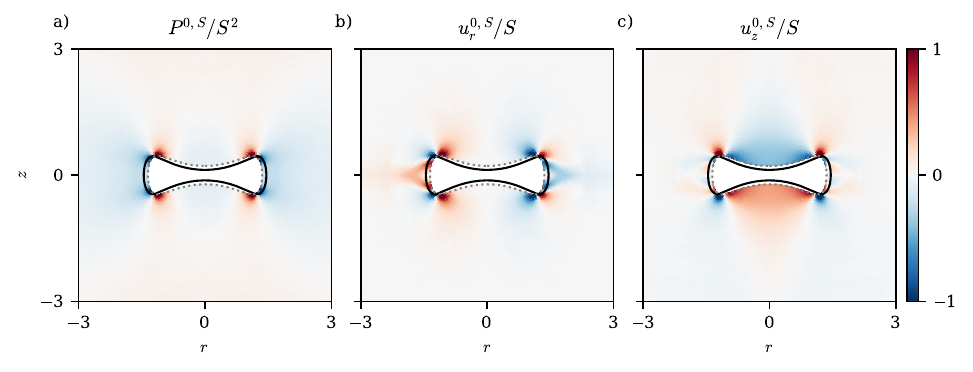}
    \caption{Some characteristics of an axisymmetric unstable (0 - S) (s) mode, here at $\Oh = 0.015$, $\We = 11.5$. (a) Pressure field; (b) radial velocity field; (c) axial velocity field. The solid black and gray dotted lines depict the disturbed and undisturbed bubble shapes, respectively (the amplitude of the disturbance is arbitrary).}
    \label{fig:EM-Unstab_0S}
\end{figure}

Figure~\ref{fig:EM-Unstab_0S} illustrates some characteristics of a typical unstable (0 - S) (s) mode, here for $\Oh = 0.015$ \textendash{} $\We =11.5$.
The bubble inflates or deflates in the vicinity of the symmetry axis, depending on the sign of the initial disturbance. The pressure and velocity fields associated with the disturbance all reach their peak values near the bubble's `edges'. Provided this  (0 - S) (s) mode dominates the dynamics and does not saturate, it could lead to breakup, forming either a toroidal bubble for $\Oh> 0.007$ (as Fig. \ref{fig:EM-Unstab_0S} prefigures) or a dumbbell bubble for $\Oh < 0.008$.

\subsection{Symmetry-breaking scenarios: Drift modes}

We now consider perturbation which allow the bubble centroid to drift, i.e. to move freely in the flow. If this drift takes place in the symmetry plane $z=0$, it breaks the axial symmetry of the problem, making the consideration of unstable modes associated with the azimuthal wavenumber $m=1$ relevant. Conversely, if the bubble drifts along the symmetry axis $r=0$, antisymmetric (A) modes come into play.
Similarly to the findings of Sierra-Ausin \textit{et al.}~\cite{sierra-Ausin2022} in a uniaxial straining flow, we identify two stationary (s) drift modes.
One arises when the bubble follows the outward flow in the symmetry plane $z=0$, while the other takes place when the bubble goes `against' the compressional flow along the symmetry axis $r=0$, in a manner resembling self-propulsion.

\begin{figure}
    \centering
    \includegraphics{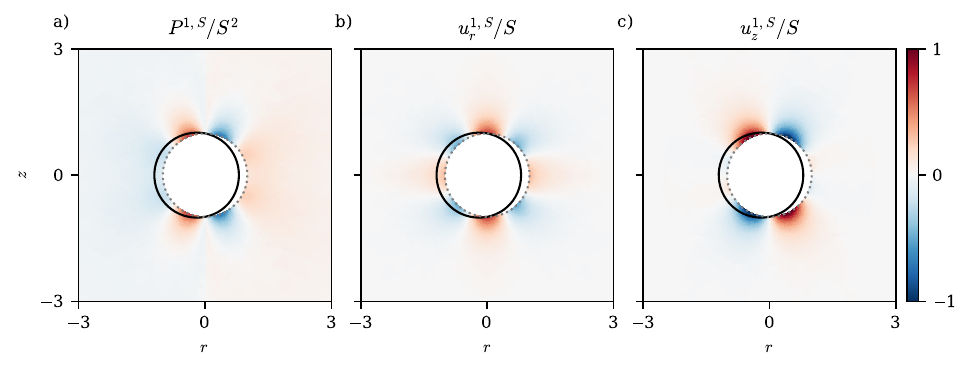}
    \includegraphics{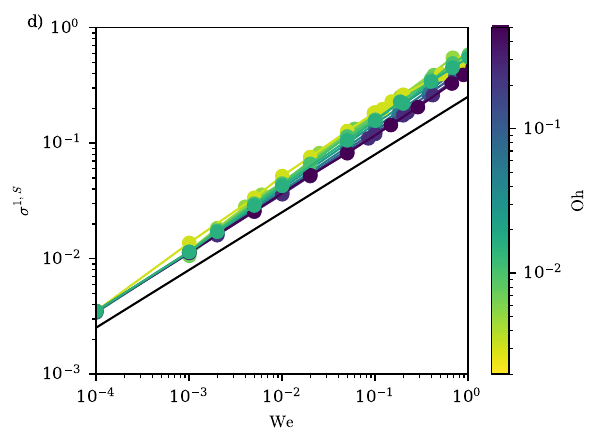}
    \caption{Some characteristics of the unstable (1 - S) (s) drift mode, here at $\Oh = 0.015$, $\We = 0.1$. (a) Pressure field; (b) radial velocity field; (c) axial velocity field. The solid black and gray dotted lines depict the disturbed and undisturbed bubble shapes, respectively (the amplitude of the disturbance is arbitrary). (d) Growth rate $\sigma^{1,S}$ of the mode as a function of $\We$, the Ohnesorge number being color-coded. The solid black line with a slope of $1/2$ is a guide to the eye. }
    \label{fig:EM-Unstab_1S}
\end{figure}

The mode associated with a drift along the outward flow exists for all pairs ($\Oh, \We$). It preserves the planar symmetry with respect to the plane $z=0$ and is stationary. Therefore, this mode is of the (1-S) (s) type.
As exemplified in Fig.~\ref{fig:EM-Unstab_1S} for $\Oh = 0.015$ \textendash{} $\We=0.1$, this mode corresponds to a bubble displacement (solid black line) in any direction lying in the plane $z=0$, with little deformation compared to the equilibrium shape (gray dotted line).
The pressure disturbance at the front of the bubble is negative while an overpressure is observed at its back, resulting in a pressure difference that contributes to the bubble drift.
Figure~\ref{fig:EM-Unstab_1S}(d) shows the growth rate $\sigma^{1, S}$ of this mode for all $\Oh$ values in the limit of small Weber numbers.
It is seen that $\sigma^{1,S}$ increases nearly linearly with $\We^{1/2}$, with only a weak dependence on $\Oh$. 
As inertial and viscous effects cooperate to displace the bubble, their relative importance has weak consequences on the bubble dynamics.
Since times are normalized by the capillary timescale $(\rho R^3/\gamma)^{1/2}$ the physical growth rate turns out to be linearly proportional to the strain rate $S$.
Therefore, the stronger the flow the faster the bubble escapes away from the stagnation point.

\begin{figure}
    \centering
    \includegraphics{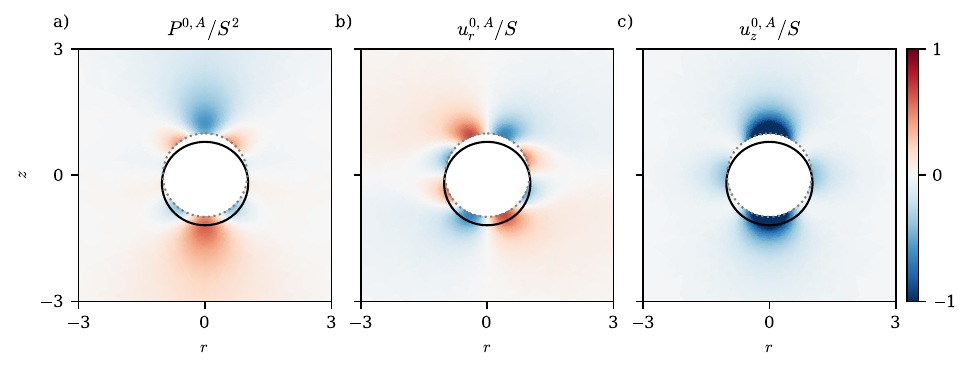}
    \includegraphics{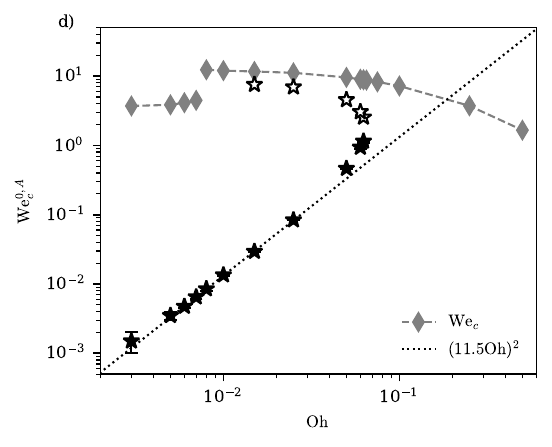}
    \includegraphics{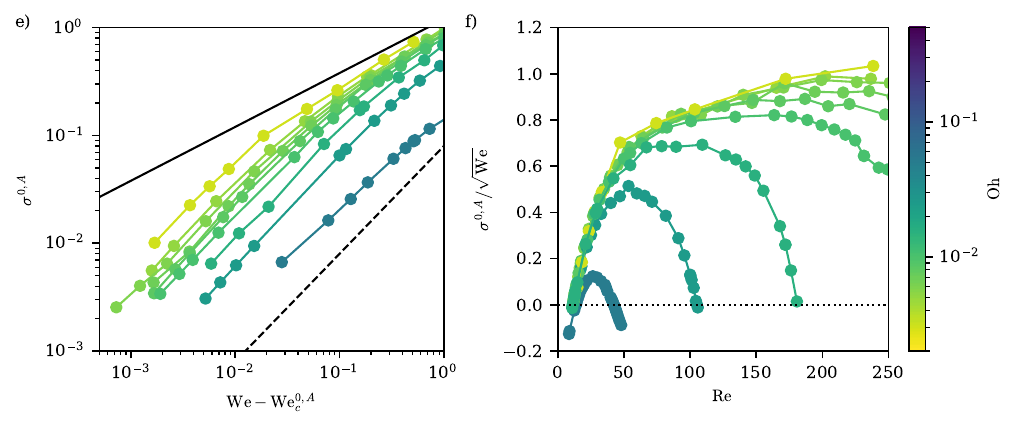}
    \caption{Some characteristics of the unstable (0 - A) (s) drift mode, here at $\Oh = 0.015$, $\We = 0.1$. (a) Pressure field; (b) radial velocity field; (c) axial velocity field. The solid black and gray dotted lines depict the disturbed and undisturbed bubble shapes, respectively (the amplitude of the disturbance is arbitrary).
    (d) Domain of existence of the mode. Solid black stars denote the critical Weber number $\We_c^{0, A}(\Oh)$ (determined by bisection), beyond which the mode becomes unstable, while white stars denote the Weber number at which it restabilizes. At low $\Oh$, the critical Weber number follows the law $\We_c^{0,A}=(11.5\, \Oh)^2$ (black dotted line). 
     The error bar on the determination of $\We_c^{0,A}$ is shown near the bottom left corner. Weber numbers $\We_{c}(\Oh)$ at the turning point (gray diamonds) represent an upper bound for $\We_c^{0, A}$. 
     (e) Growth rate $\sigma^{0, A}$ as a function of the distance to the critical Weber number; the Ohnesorge number is color-coded. The solid and dashed black lines with slopes 1/2 and 1 are guides to the eye.
     (f)~Normalized growth rate $\sigma^{0, A}\We^{-1/2}$ as a function of the Reynolds number $\Rey=\We^{1/2}\Oh^{-1}$.}
    \label{fig:EM-Unstab_0AS}
\end{figure}

The second drift mode, associated with a bubble displacement `against' the flow along the symmetry axis is also stationary, making it of the (0-A) (s) type.
Figure~\ref{fig:EM-Unstab_0AS}(a-c) shows some characteristics of this mode for $\Oh = 0.015$ and $\We = 0.1$. 
In contrast to the above (1-S) (s) mode, this mode becomes unstable beyond a finite critical Weber number $\We_c^{0, A}(\Oh)$, and restabilizes for large enough $\We$, as the black and white stars in Fig.~\ref{fig:stab}(d) indicate.
The critical Weber number $\We_c^{0, A}$ increases sharply as viscous effects become more prominent, following the approximate law $\We_c^{0, A} \approx (11.5\, \Oh)^2$ in the limit of small $\Oh$.
Remembering that $\We = (\Rey \, \Oh)^2$, it follows that the (0 - A) (s) mode becomes unstable beyond a critical Reynolds number $\Rey_c^{0, A} \approx 11.5$ when $\Oh$ is small enough.
Note that a similar result was found in the uniaxial configuration, the critical Reynolds number being around 20 in that case \cite{sierra-Ausin2022}.
Obviously, the largest $\We$ at which the (0 - A) (s) mode may exist cannot exceed the critical Weber number at the turning point, $\We_{c}(\Oh)$. Since $\We_{c}(\Oh)$ decreases with increasing $\Oh$ according to Fig. \ref{fig:phase diagram}(a), one can anticipate the existence of a maximum Ohnesorge number beyond which the (0 - A) (s) mode disappears. This is confirmed by Fig. \ref{fig:EM-Unstab_0AS}(d) which reveals that this mode exists only up to $\Oh\approx 0.065$. As discussed in \cite{sierra-Ausin2022}, this mode  is primarily governed by pressure effects. Indeed, keeping in mind that $P_{BSF} = -S^2 (\frac{1}{8}r^2 + \frac{1}{2}z^2)$, it appears that, at every point of the flow, the pressure gradient in the base flow is oriented towards the stagnation point, so that pressure effects drive the bubble away from that point as soon as its centroid departs from the origin $r=z=0$. Therefore,   
when viscous effects are weak enough, i.e., $\Rey$ is sufficiently large, they cannot prevent the bubble from following the pressure loss along the symmetry axis, even though the flow itself is locally oriented towards the stagnation point.
Figure~\ref{fig:EM-Unstab_0AS}(e) shows the growth rate $\sigma^{0, A}$ of the (0 - A) (s) mode as a function of the distance to the critical Weber number, $\We - \We_c^{0, A}$.
In contrast with the growth rate of the (1 - S) (s) mode, $\sigma^{0, A}$ strongly depends on $\Oh$: at a distance $\We - \We_c^{0, A}=10^{-2}$, it varies by more than one order of magnitude, from $1.2\cdot 10^{-3}$ at $\Oh=0.05$ to $4\cdot 10^{-2}$ at $\Oh=0.003$. 
Therefore, it turns out that the stability of this mode is controlled by the balance between inertial and viscous effects (synthesized in  $\Rey_c^{0, A}$) while its growth rate is controlled by the balance between capillary and viscous effects (synthesized in  $\Oh$).
Close to $\We_c^{0, A}$, the growth rate increases linearly with the distance $\We-\We_c^{0, A}$.
However, $\We_c^{0, A}$ goes to zero as $\Oh$ decreases. In that limit, $\sigma^{0, A}$ is seen to grow like $\We^{1/2}$ for Weber numbers of order unity. Therefore, the advective limit already discussed in connection with the (1 - S) (s) mode is recovered under such conditions. This limit is better visualized in Fig.~\ref{fig:EM-Unstab_0AS}(f) which shows the rescaled growth rate $\sigma^{0, A}\We^{-1/2}$ as a function of the Reynolds number $\Rey=\We^{1/2}\Oh^{-1}$. In the limit of large Reynolds number, the rescaled growth rate converges to a plateau, confirming that $\sigma^{0, A}$ grows linearly with the strain rate $S$. The figure also confirms the magnitude of the critical Reynolds number, $\Rey_c^{0, A}\approx11.5$. For the largest three $\Oh$, it also highlights the upper Reynolds number at which the mode (0-A) (s) re-stabilizes. 

\subsection{Higher-order modes}
\begin{figure}
    \centering
    \includegraphics{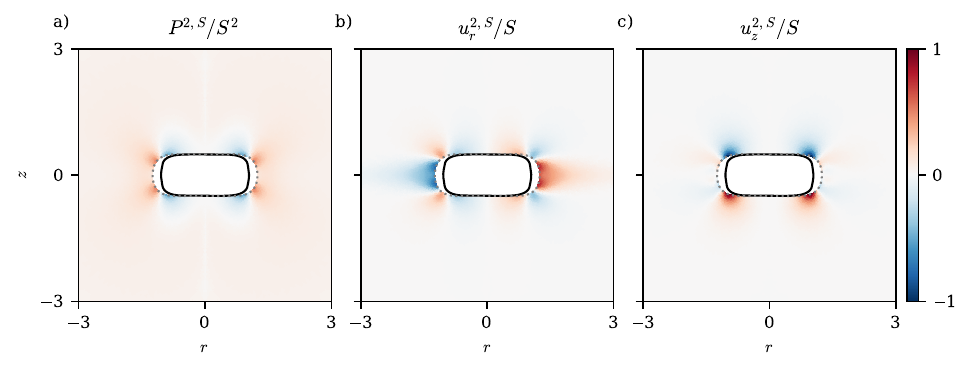}
    \includegraphics{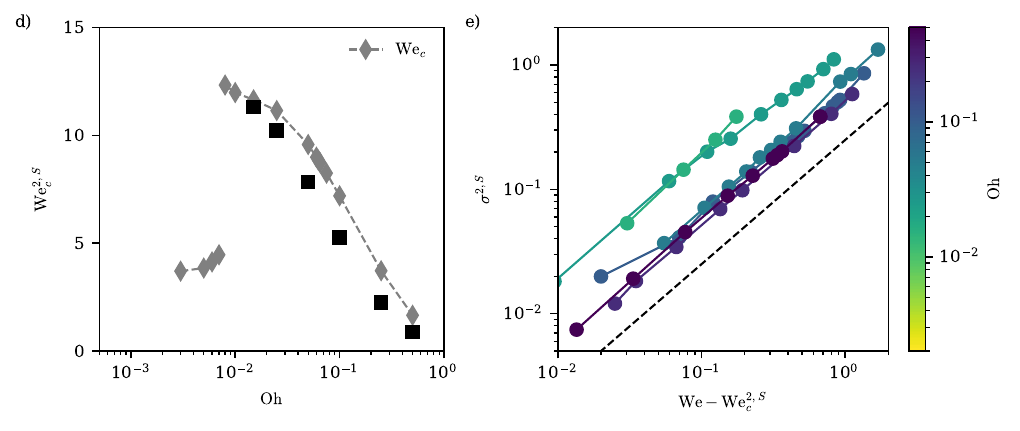}
    \caption{
    Some characteristics of the unstable (2 - S) (s) mode, here at $\Oh = 0.05$, $\We = 8.2$. (a) Pressure field; (b) radial velocity field; (c) axial velocity field. The solid black and gray dotted lines depict the disturbed and undisturbed bubble shapes, respectively (the amplitude of the disturbance is arbitrary). 
    (d) Critical Weber number $\We_c^{2, S}$ as a function of the Ohnesorge number (black squares), found by bisection. 
    The Weber number $\We_{c}(\Oh)$ at the turning point (gray diamonds) being an upper bound for $\We_c^{2, S}$, there is a minimum Ohnesorge number, $\Oh \approx 0.015 $, below which the mode disappears. 
    (e) Variation of the growth rate $\sigma^{2,S}$ with the distance $\We-\We_c^{2,S}$; the Ohnesorge number is color-coded. The dashed black line with slope 1 is a guide to the eye. 
    }
    \label{fig:EM-Unstab_2S}
\end{figure}

In uniaxial straining flows, no modes with azimuthal wavenumbers $m>1$ have been reported. This is because the large radial curvatures associated with the prolate equilibrium shapes prevent the occurrence of large deformations within the symmetry plane.
In biaxial flows, however, equilibrium shapes are frequently oblate, as Fig. \ref{fig:phase diagram}(a) revealed. Therefore, one can expect higher-order modes to arise on such bubbles.
Indeed, on branches that are stable with respect to axisymmetric disturbances, we found that stationary (2 -S) (s) modes may exist beyond a critical Weber number $\We_c^{2, S}(\Oh)$.
An example of such a mode is given in Fig.~\ref{fig:EM-Unstab_2S} for $\Oh = 0.05$ and $\We=8.2$.
These modes give rise to deformations of the disk-like base shapes into flat peanuts: while a compression is observed in the diametrical plane represented in Fig.~\ref{fig:EM-Unstab_2S}(a-c) (which corresponds to azimuthal angles $\theta=0,\pi$), an extension occurs in the orthogonal diametrical plane corresponding to $\theta=\pm\pi/2$.
Obviously, the orientation of the compressional and extensional directions is arbitrary, being set by the initial disturbance.
The critical Weber number $\We_c^{2, S}(\Oh)$ beyond which this mode destabilizes decreases with increasing $\Oh$, as shown in Fig.~\ref{fig:EM-Unstab_2S}(d).
Due to the variations of the critical Weber number $\We_c(\Oh)$ with $\Oh$, there is a minimum Ohnesorge number, $\Oh\approx 0.015$ below which no (2 - S) (s) mode is observed.
As shown in Fig.~\ref{fig:EM-Unstab_2S}(e), the growth rate $\sigma^{2,S}$ increases linearly with the distance $\We-\We_c^{2,S}$ for all investigated $\Oh$.
The pre-factor remains of order unity when varying $\Oh$ by a factor of 25 in the range $0.02 - 0.5$.
In physical units, this implies that the growth rate is proportional to $(\rho R^3/\gamma)^{1/2}(S^2-S_c^{2})$, with $S_c$ the critical strain rate beyond which the mode destabilizes.
This scaling law indicates that the instability process is governed by capillarity and strain. More precisely, what happens is that, for sufficiently flat disk-like bubble shapes, the radial curvature becomes so low that capillary effects are unable to counterbalance the destabilizing effect of the ambient strain, leading to an instability. 
Importantly, as Fig.~\ref{fig:EM-Unstab_2S}(d) makes clear, the (2 - S) (s) mode arises at Weber numbers lower than $\We_c$, i.e., slightly before the turning point is reached and the axisymmetric (0 - S) (s) mode becomes unstable. 
Provided  the (2 - S) (s) modes do not saturate due to nonlinear effects and `win' against the drift modes, they should lead to bubble breakup following a scenario qualitatively similar to that observed in the uniaxial configuration. An experimental evidence of such breakup scenario might be described in \cite{ravelet2011} for bubbles rising in turbulence, where oblate bubbles break in two, following non oscillating deformations.
We did not observe any (2 - A) mode, nor higher-order modes ($m>2$) with either symmetry.
We hypothesize that higher-order modes require even flatter oblate shapes that are not encountered in the presence of finite inertia effects.

\section{Summary and prospects}

\begin{figure}
    \centering
    \includegraphics{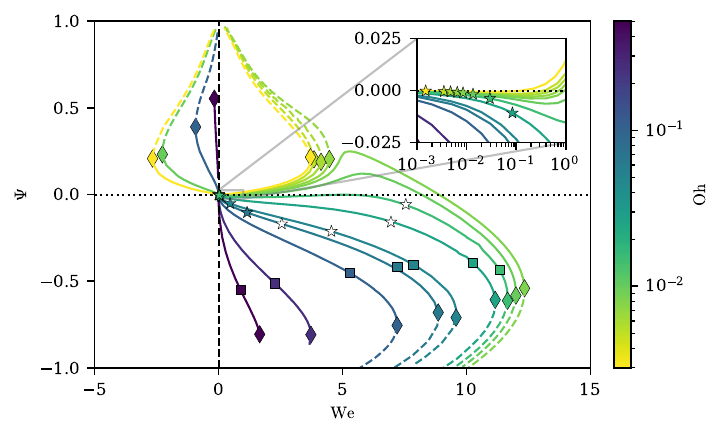}
    \caption{Summary of the findings obtained through the L-ALE approach in both biaxial and uniaxial straining flows. Curves on the $\We>0$ side reproduce those of figure \ref{fig:phase diagram}(a) for the biaxial case. Results for the uniaxial case are displayed on the $\We<0$ side for convenience; they reproduce those of figure (2a) of Ref~\cite{sierra-Ausin2022}, using parameter $\Psi$ instead of the aspect ratio $\chi$ to characterize the shape.
    Unstable axisymmetric (0 - S) (s) modes occur on the folded branches, whatever $\Oh$ (dashed lines), starting at the critical Weber number, $\We_c(\Oh)$ (diamonds).
    Asymmetric (1 - S) (s) modes corresponding to a drift along the extensional $z=0$ plane are present whatever $\We$. 
    Unstable (0 - A) (s) modes corresponding to a drift `against' the flow along the compressional axis $r=0$ occur on primary branches at the position of the colored stars and disappear at that of white stars. They remain stable whatever $\We$ for the largest two $\Oh$. 
    Modes (2 - S) (s) occur for $\Psi \approx -0.5 $  (squares) on the primary branches at locations where axisymmetric 0 - S modes are still stable. Inset: enlargement in the vicinity of $\Psi=0$, showing the variation of the critical Weber number $\We_c^{0, A}$ of the (0 - A) (s) mode with $\Oh$. The four branches reported in the quadrant $\We<0, \Psi>0$ correspond to the equilibrium positions of bubbles held fixed at the stagnation point of a uniaxial straining flow.}
    \label{fig:stab}
\end{figure}

In this paper, we made use of the recently developed L-ALE approach to investigate the rich dynamics of an incompressible gas bubble immersed at the stagnation point of a biaxial straining flow. The results of this investigation are summarized in Fig.~\ref{fig:stab}. For the sake of broader comparison, this figure also shows four equilibrium branches of bubbles in a uniaxial straining flow, corresponding to $\Oh \in [1.10^{-3}, 0.5]$, obtained with the same code and encoded in the half-plane $\We\leq 0$ for better readability.
In a biaxial flow, regardless of $\Oh$, we found that the system exhibits a saddle-node bifurcation, similar to what is observed in uniaxial straining flows. Yet, strong qualitative differences arise when $\Oh$ decreases. Over a significant range of Weber number, bubble shapes transition from oblate spheroids to (counter-intuitive) prolate spheroids, owing to dominant pressure forces. 
In addition, two new families of disconnected branches have been uncovered. While hourglass bubbles are largely unstable and can presumably not be observed in practice, the second set of S-shaped branches exhibits a stable part with respect to axisymmetric disturbances and might be observable.
By investigating the linear stability of the various branches, we evidenced that most of the phenomenology at play is qualitatively similar to what is observed in uniaxial straining flows. That is: (i) unstable stationary axisymmetric modes start growing at the turning point; (ii) an unstable drift mode exists whatever $\We$ along any radial direction; and (iii) a second, counter-intuitive, drift mode along the axial direction is unstable within a finite range of $\We$.
However, for large enough $\Oh$, we also uncover a new unstable mode with higher azimuthal wavenumber, $m =2$.
This mode exists on oblate equilibrium bubble shapes that are still stable with respect to axisymmetric disturbances. 
Interestingly, Fig.~\ref{fig:stab} reveals that the shape factor $\Psi$ at the critical Weber number $\We^{2,S}$ beyond which this mode is unstable only loosely depends on $\Oh$ and is close to $\Psi \approx -0.5$. This suggests that the (2 -S) (s) mode occurs when the bubble oblateness is large enough, meaning when the axial curvature of the interface is small enough. Hence, the reason why $\We_c^{2,S}$ decreases with increasing $\Oh$, as the figure makes clear, is not because viscous effects enhance the instability underlying this mode, but rather because they enhance the bubble oblateness at a given Weber number. Provided this $m=2$ mode does not saturate and overwhelms the drift modes, it can lead to bubble breakup.
In this scenario, the breakup sequence would be largely similar to what is observed in uniaxial straining flows. It might explain some of the breakage events observed in real configurations under subcritical conditions.

Some of the results obtained in this investigation may be of interest in the context of bubble deformation and breakup in turbulence.
First, prolate shapes can be observed in both uniaxial and biaxial straining flows, provided $\Oh$ is small enough. While the axisymmetric mode is qualitatively similar in both configurations, it would be of great interest to determine what is the dominant flow geometry triggering breakup, as this geometry might influence the forthcoming fragmentation.
Second,  the $\We$-range where bubbles are stable in the biaxial configuration is much larger than in the uniaxial one where the critical Weber number $\We_c$ is less than $3.0$, even for vanishingly small $\Oh$ \cite{miksis1981, sierra-Ausin2022} (see the corresponding branches reported in the top left part of Fig. \ref{fig:stab}). This stable range, which extends beyond $\We=10$ when viscous effects are weak enough, may be even wider when accounting for presumably stable shapes lying on disconnected S-shaped branches. Since the biaxial geometry is the most probable local configuration in isotropic turbulence, the above findings are of importance in the context of bubble transport and breakup in turbulence. On this basis, we hypothesize that the dominant behavior of bubbles in biaxial straining regions is a drift away from the stagnation point in the plane where the main extension takes place.
Nonetheless, we cannot exclude that, on the way to escape, nonlinear couplings between drift modes and deformation modes may lead to breakup. Such couplings remain to be studied.

\begin{acknowledgments}
The first author acknowledges the support from the Swiss National Foundation through the SNSF Swiss Postdoctoral Fellowship, project number 233920.
\end{acknowledgments}

\appendix
\section{Comparison of the equilibrium shapes with the results of \cite{kang1989biaxial}}
\begin{figure}
    \centering
    \includegraphics{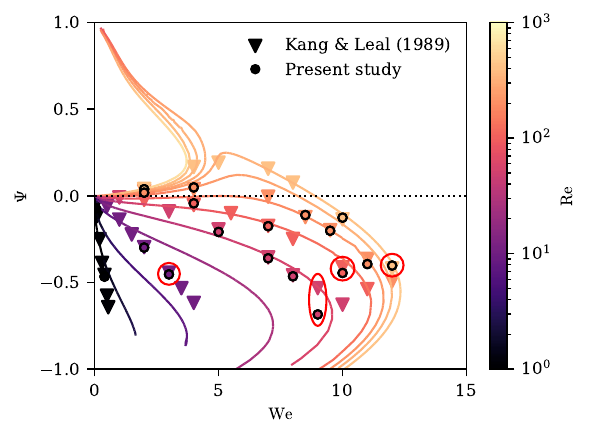}
    \caption{Comparison of equilibrium shapes with those of \cite{kang1989biaxial}. Solid lines color-coded with the Reynolds number: data computed here at fixed $\Oh$ (see Sec. \ref{sec:eqpos}). Triangles (color-coded similarly): data from \cite{kang1989biaxial}, extracted from bubble shapes given in figure 4 of that reference; black circles: additional simulations performed with the L-ALE approach for some specific parameter sets considered in \cite{kang1989biaxial}. Red circles: data sets for which the mismatch between present results and predictions from \cite{kang1989biaxial} is the largest; the corresponding data are reported in Table \ref{tab:compKL}.  }
    \label{fig:compKL}
\end{figure}

Kang and Leal (Ref. \cite{kang1989biaxial}) investigated the stable equilibrium shapes of gas bubbles held fixed at the stagnation point of a biaxial flow at eight different Reynolds numbers, including $\Rey = 0$ and $\Rey = \infty$ (potential flow).
Figure~\ref{fig:compKL} compares the equilibrium shape factor $\Psi$ obtained with our L-ALE approach with these earlier results obtained with a time-marching finite-difference approach using a $40\times40$ spatial resolution. Solid lines are the continuations at fixed Ohnesorge number which are discussed in Sec. \ref{sec:eqpos}, here color-coded by the Reynolds number, which is not constant along a given branch. 
Triangles show the equilibrium positions found in \cite{kang1989biaxial} ($\Psi$-values were determined from bubble shapes displayed in figure 4 of that reference). Circular markers represent additional L-ALE simulations matching some parameters sets of \cite{kang1989biaxial}. As long as $\Psi > -0.5$, the two datasets are in excellent agreement.
However, close to the turning point, where equilibrium positions strongly depend on $\We$, our solutions consistently exhibit larger deformations than those of \cite{kang1989biaxial}.
The two sets of results at the four points circled in red in the figure, for which the mismatch is the most severe, are reported in Table~\ref{tab:compKL}. 

\begin{table}[]
    \centering
    \begin{tabular}{c|c|c}
         $(\Rey, \We$)& $\Psi$ (Ref. \cite{kang1989biaxial}) & $\Psi$ (Present study)  \\
         \hline
         (10, 3)& -0.44 & -0.452\\
         (50, 9) & -0.46 &-0.684 \\
         (100, 10 ) & -0.41 &-0.445 \\
         (400, 12) & -0.42 & -0.402
    \end{tabular}
    \caption{Comparison between present predictions and results of \cite{kang1989biaxial} for the four sets of parameters circled in red in Fig.~\ref{fig:compKL} for which the mismatch is the most severe. }
    \label{tab:compKL}
\end{table}
\bibliography{biblio}

\end{document}